\documentclass[preprint,authoryear,11pt]{elsarticle}

\usepackage{graphicx}
\usepackage{amssymb}

\usepackage{arydshln} 
\usepackage{multirow}
\usepackage{subfig} 
\usepackage{amsmath, bm}
\usepackage{booktabs}
\usepackage{soul, color} 
\usepackage{arydshln} 
\usepackage{hyperref}
\hypersetup{colorlinks,allcolors=black}

\DeclareMathOperator*{\argmin}{argmin} 

\biboptions{numbers,compress}

\journal{Artificial Intelligence in Medicine}

\begin{document}

\begin{frontmatter}


\title{Enhancing Fiber Orientation Distributions using Convolutional Neural Networks}




\author[bmeis]{Oeslle Lucena\corref{mycorrespondingauthor}}
\cortext[mycorrespondingauthor]{Corresponding author}
\ead{oeslle.lucena@kcl.ac.uk}
\author[cmic,queen]{Sjoerd B. Vos}
\author[dept_epilepsy_ucl]{Vejay Vakharia}
\author[dept_epilepsy_ucl,nhs_qs]{John Duncan}
\author[kch]{Keyoumars Ashkan}
\author[bmeis]{Rachel Sparks}
\author[bmeis]{Sebastien Ourselin}

\address[bmeis]{School of Biomedical Engineering and Imaging Sciences, King's College London, UK}
\address[cmic]{Centre for Medical Image Computing, Department of Computer Sciences, University College London, London, UK}
\address[queen]{Neuroradiological Academic Unit, University College London Queen Square Institute of Neurology, University College London, London, UK}
\address[soc_epilepsy]{Epilepsy Society MRI Unit, Chalfont St Peter, UK}
\address[dept_epilepsy_ucl]{Department of Clinical and Experimental Epilepsy, University College London, UK}
\address[nhs_qs]{National Hospital for Neurology and Neurosurgery, Queen Square, UK}
\address[kch]{King's College Hospital Foundation Trust, UK}

\begin{abstract}
Accurate local fiber orientation distribution (FOD) modeling based on diffusion magnetic resonance imaging (dMRI) capable of resolving complex fiber configurations benefits from specific acquisition protocols that sample a high number of gradient directions (b-vecs), a high maximum b-value (b-vals), and multiple b-values (multi-shell). However, acquisition time is limited in a clinical setting and commercial scanners may not provide such dMRI sequences. Therefore, dMRI is often acquired as single-shell (single b-value). In this work, we learn improved FODs for commercially acquired dMRI. We evaluate patch-based 3D convolutional neural networks (CNNs) on their ability to regress multi-shell FOD representations from single-shell representations, where the representation is a spherical harmonics obtained from constrained spherical deconvolution (CSD) to model FODs. We evaluate U-Net and HighResNet 3D CNN architectures on data from the Human Connectome Project and an in-house dataset. We evaluate how well each CNN model can resolve local fiber orientation 1) when training and testing on datasets with the same dMRI acquisition protocol;
2) when testing on a dataset with a different dMRI acquisition protocol than used to train the CNN models; and 3) when testing on a dataset with a fewer number of gradient directions than used to train the CNN models. Our approach may enable robust CSD model estimation on single-shell dMRI acquisition protocols with few gradient directions, reducing acquisition times, facilitating translation of improved FOD estimation to time-limited clinical environments.
\end{abstract}


\end{frontmatter}


\label{sec:Introduction}
\section{Introduction}
Diffusion MRI (dMRI) captures water molecule diffusion and can reveal the underlying organization of aspects relating to different tissue components. In the brain, dMRI is used to investigate the organization of the white matter (WM), which is composed of neuronal axon bundles that impose a direction to the diffusion of water molecules~\cite{tanner1979self, shapey2019clinical}, resulting in anisotropic diffusion with the preferred direction of diffusion being along the axon fiber~\cite{basser1994mr, jeurissen2014multi}.  

From acquired dMRI signals, it is possible to non-invasively estimate WM tissue microstructure information, such as axon diameter~\cite{assaf2008axcaliber}, and compute local fiber orientation distribution (FOD or fODF) at the voxel level~\cite{alexander2002detection}. FODs can be used as input to perform fiber tractography~\cite{berman2009diffusion,jeurissen2019diffusion} which has an important role in presurgical planning~\cite{winston2014preventing, essayed2017white, mancini2019automated, o2017automated} and connectome analyses~\cite{setsompop2013pushing}. A common method to estimate local fiber orientation is diffusion tensor imaging (DTI)~\cite{basser1994mr}. However, DTI can only model a single fiber population and it cannot resolve complex fiber configurations in the brain such as crossing fibers~\cite{alexander2002detection}. To address this issue, more robust methods for representing FODs have been presented that can resolve fiber crossing based on spherical deconvolution~\cite{canales2019sparse, dell2018modelling} or other approaches that estimates diffusion orientation distribution functions from q-space~\cite{westin2016q,dell2018modelling}.

Constrained spherical deconvolution (CSD) is one method capable of estimating a FOD from dMRI signals~\cite{tournier2007robust,jeurissen2014multi}. Single-shell single-tissue CSD (S-CSD) models each voxel as a single compartment with one corresponding FOD, irrespective of any underlying tissue components~\cite{tournier2007robust}. However, S-CSD FODs suffers from partial volume effects (PVE) when multiple tissues are present in a voxel. Multi-shell multi-tissue CSD (M-CSD) extends S-CSD by modeling one anisotropic compartment (corresponding to WM) and two isotropic compartments (corresponding to GM and CSF), which provides a more reliable and accurate estimation of the WM FOD~\cite{jeurissen2014multi}. M-CSD makes use of each tissue type's distinct decay behaviour across multiple b-values (or shells) to separate each voxel into three components. Nevertheless, a limitation with M-CSD is that it requires multi-shell (MS) dMRI which has longer acquisition times compared to DTI or SS dMRI. 

Alternatively, single-shell 2-tissue (2TS-CSD) and single-shell 3-tissue CSD (SS3T-CSD)~\cite{dhollander2016novel} attempt to resolve the PVE by modeling isotropic compartments, similar to M-CSD, using the b=$0$ s/mm$^2$ scan as a second shell. With the additional shell, a multi-tissue signal profile is computed by 2TS-CSD assuming one anisotropic component (WM) and one isotropic component (either GM or CSF). Compared to M-CSD, 2TS-CSD can only model one isotropic tissue. SS3T-CSD uses an iterative approach to fit a CSD model for the three tissues compartments by first fitting one pair of components (anisotropic and one isotropic) and then fitting another pair of components (anisotropic and one isotropic)~\cite{dhollander2016novel}.


Usually, accurate FOD models that are able to resolve complex fiber configurations require
specific dMRI acquisition protocols with a high number of gradient directions~(b-vecs), a high maximum b-value and/or multiple b-values~\cite{neher2017fiber, daducci2013quantitative,descoteaux1999high,vos2016trade}. A higher number of gradient directions and high b-values improve FOD angular resolution, enabling FODs to better resolve complex fiber configurations, such as fiber crossings~\cite{jones2013white,tournier2013determination}. Multiple b-values allow for multiple compartment modeling, correcting for PVE~\cite{jeurissen2014multi}. Additionally, local fiber reconstruction is more accurate for images with a high signal-to-noise ratio (SNR), which can be achieved from low b-value shells~\cite{tournier2013determination}. All of these constraints for dMRI acquisition impose longer acquisition times. Despite these advantages, clinical uptake has been limited due to the longer acquisition times and the need for expert staff to set up the acquisition protocols~\cite{ordonez2019identification}. Therefore, improving FOD modeling for commercially available dMRI acquisitions is an active topic of research. 

Deep learning (DL) has been successfully implemented for a variety of medical imaging tasks~\cite{litjens2017survey}. DL methods learn an underlying mathematical representation that can non-linearly map data from one representation to another representation, for instance mapping from raw image intensity to a predicted class or another intensity space~\cite{lecun2015deep}. DL has been successfully applied to dMRI through image quality transfer (IQT) to improve its spatial resolution~\cite{alexander2017image,TannoBayesIQT2017}, fit neurite orientation dispersion and density imaging (NODDI)~\cite{zhang2012noddi} and spherical mean technique (SMT)~\cite{kaden2016quantitative} maps from q-space~\cite{alexander2017image,golkov2016q}, predict high order of CSD model coefficients from CSD model coefficients computed on downsampled dMRI~\cite{lin2019fast}, and improve FOD model fitting~\cite{koppers2016diffusion,nath2019deep,lin2019fast}. 

In~\cite{alexander2017image}, IQT was proposed with linear regression and random forest models to (a) infer high-resolution dMRI patches from a lower spatial resolution and (b) to learn a mapping between parameters of different models. dMRI parameter mapping was evaluated to go from a low order DTI model to a higher order model; either NODDI or SMT were both evaluated. Further work evaluated IQT using a CNN patch-based regression to infer higher resolution patches and quantified regression uncertainty~\cite{TannoBayesIQT2017}. Regression between models has been also presented for diffusion kurtosis imaging~\cite{lu2006three} and NODDI from shorter q-space MS dMRI signal intensities using a multilayer perceptron network (MLP)~\cite{golkov2016q}.

In~\cite{koppers2016diffusion}, a MLP was trained to infer SH coefficients across different shells. Here, a MLP used SH model coefficients of the same order calculated from one shell or combination of shells to infer the SH coefficients for a different shell. The MLP was restricted to use shells with the same number of gradient directions. One limitation of this approach is that the mapping only used voxel information and did not take into account neighborhood information, which may provide important spatial context. Another limitation is that this approach did not evaluate if the method would generalize to other acquisition protocols. 

In~\cite{nath2019deep}, a neural network composed of regular hidden and residual layers (ResDNN) was trained to map S-CSD coefficients to FODs computed from histology. The angular correlation coefficient (ACC) was used to evaluate FOD accuracy using FODs computed from histology as ground truth. ResDNN outperformed other methods to estimate FODs including S-CSD and Q-ball imaging. ResDNN was also evaluated for reproducibility on 12 paired in-vivo dMRI obtained from the human connectome project (HCP). 
The major disadvantage of this work is that the model is trained on data acquired from Macaque imagery and then transferred to human dMRI. As there is no ground truth histology FOD for the human brain, it was not clear whether ResDNN improves the FOD estimation on this dataset. Furthermore, the baseline comparison used for the human imaging was S-CSD which is suboptimal as several more robust approaches exist - e.g. M-CSD.

In~\cite{lin2019fast}, 3D CNNs were used to estimate a CSD model coefficients, either M-CSD or S-CSD coefficients, from CSD models coefficients computed on downsampled dMRI -- gradients directions were downsampled compared to original dMRI. In this work, only fitting between the same type of CSD model was performed (S-CSD (downsampled) to S-CSD (full) and M-CSD (downsampled) to M-CSD (full)). The 3D CNN CSD coefficients are more similar to the the "ground truth" CSD model coefficients (full) compared to the downsampled CSD model coefficients.  
In this work, the authors focused in improving CSD model fitting with fewer gradient directions but with the same number of shells , i.e. S-CSD to S-CSD or M-CSD to M-CSD. In our work, we are interested in regress CSD model coefficients from one shell (S-CSD) to multiple shells (M-CSD).

In this work, we aim to compute a more accurate and reliable FOD using data that are still the most common in clinical settings: single-shell dMRI. To achieve this, we present a framework to train a CNN to regress M-CSD model coefficients from 2TS-CSD model coefficients using a patch-based approach. We used two different 3D CNN architectures and evaluate the CNN models on two datasets. The models undergo an extensive evaluation involving: (1) training and testing on datasets with the same dMRI acquisition protocol, (2) testing on datasets with different dMRI acquisition protocol than the training dataset, and (3) testing on dMRI that have fewer gradient directions.


\label{sec:Methods}
\section{Methodology}
\subsection{Pipeline Overview}
Our pipeline consists of the following steps.
We use data from the Human Connectome Project~(HCP)~\cite{sotiropoulos2013advances} and an in-house dataset which we refer to as QS dataset~(see dataset and preprocessing details in Section~\ref{sec:datasets}). Secondly, we construct a paired dataset composed of SS dMRI and the original MS dMRI~(Section~\ref{sec:train_data}). From this data, we compute two CSD models 2TS-CSD and M-CSD~(Section~\ref{sec:csd_modeling}) from the SS dMRI and MS dMRI data, respectively.
Finally, we train a CNN model to regress the M-CSD model coefficients from the 2TS-CSD model coefficients~(Section~\ref{sec:csd_modeling}) using the paired dataset.   

\subsection{Training Dataset}
\label{sec:train_data}
From the preprocessed MS dMRI, a paired SS dMRI is constructed by selecting an appropriate shell from the MS dMRI. In this work, we select a shell based on the compromise of a minimum number of gradient directions for a given b-value that can best characterize the angular frequency components of the dMRI signal~\cite{tournier2013determination}.

For the HCP dataset we construct SS dMRI for all $3$ b-values ($1000$, $2000$, $3000$ s/mm$^2$) where each shell has $90$ directions. For the QS we construct paired SS dMRI for $2$ b-values ($700$, $2500$ s/mm$^2$) with $32$ and $64$ directions for each shell, respectively.

\subsection{CSD modeling}
\label{sec:csd_modeling}
CSD models the FOD as a of coefficients in SH domain and applies a non‐negativity constraint as a soft regularizer using an iterative linear least-squares fit~\cite{tournier2007robust}. The original dMRI signal intensities are approximated by a convolution of the FOD model with a signal attenuation profile of a single fiber population, referred to as response function~\cite{tournier2007robust}. 

Although $l_{max} = 8$ can provide an FOD with a higher angular contrast for high b-values ($b > 1000$ s/mm$^2$)~\cite{tournier2013determination}, we focused on $l_{max} = 4$, comprising $15$ coefficients, as the order of our CSD modeling to keep the problem sufficiently simple to establish a proof of concept. 

After computing the CSD model coefficients, we applied a multi-tissue informed log-domain intensity normalization~\cite{raffelt2017bias} to both the M-CSD and the 2TS-CSD coefficients to correct intensity inhomogeneities.

\subsubsection{M-CSD Model}
M-CSD is computed using a constrained least squares fit of the dMRI signal intensities following the equation~\cite{jeurissen2014multi}:

\begin{equation}
\label{eq:csd_fiting}
    \begin{aligned}
 \begin{bmatrix} 
 {\bm {\hat x}_{1}}
 \\\vdots
 \\ {\bm {\hat x}_{n}}
 \end{bmatrix}
 =
\argmin_{{\bm x}}
\left\|
\begin{bmatrix}
{\bm{C}_{1,1}} & \cdots & {\bm{C}_{1,n}} \\
\vdots & \ddots & \vdots\\
{\bm{C}_{m,1}} & \cdots & {\bm{C}_{m,n}}
\end{bmatrix}
\begin{bmatrix} 
{\bm {x}_{1}}\\
\vdots\\ 
{\bm {x}_{n}}
\end{bmatrix}
-
\begin{bmatrix} 
{\bm {d}_{1}}\\
\vdots\\ 
{\bm {d}_{m}}
\end{bmatrix}
\right\|_{2}^{2}
\\ 
\quad \quad
\text{subject to} 
\begin{bmatrix}
\bm{A}_{1} & 0 & 0 \\
0 & \ddots & 0\\
0 & 0 & \bm{A}_{n}
\end{bmatrix}
\begin{bmatrix} 
\bm {x}_{1}\\
\vdots\\ 
\bm{x}_{n} 
\end{bmatrix}
\geq {\bm 0} 
    \end{aligned}
\end{equation}

where $\bm{d}_{i}$ is the vector of dMRI signal intensities on the $i$-th shell, $\bm{x}_{j}$ is an unknown vector of coefficients for the FOD of tissue $j$, $\bm{C}_{i,j}$ is the matrix relating the coefficients of the FOD of tissue $j$ to the dMRI signal intensities measured on the $i$-th shell in q-space by spherical convolution. An additional constraint is imposed on the coefficients fit where $\bm{A}_{j}$ is a tissue specific matrix relating the coefficients of the FOD for tissue $j$ to their signal amplitudes, effectively imposing positivity on all coefficients. To perform the optimization, we use the original optimization algorithm available within MRtrix~\cite{tournier2019mrtrix3}.

\subsubsection{2TS-CSD Model}
For the 2TS-CSD model, a similar approach to the M-CSD model is used where the b = $0$ s/mm$^2$ (b-zero) image is used as a second shell. To ensure Equation~\ref{eq:csd_fiting} has a unique solution, we set $j = 2$, reducing the number of tissue components to two. We model one compartment as isotropic corresponding to CSF, and the other compartment as anisotropic, corresponding to WM. The CSF is selected as the isotropic compartment as it leads to a more accurate fit of the FOD compared to using GM as the isotropic compartment~\cite{dhollander2016novel}.

\subsection{CNN training}
\label{sec:cnn_training}
\begin{figure*}[!htb]
\begin{center}
\includegraphics[width=\textwidth]{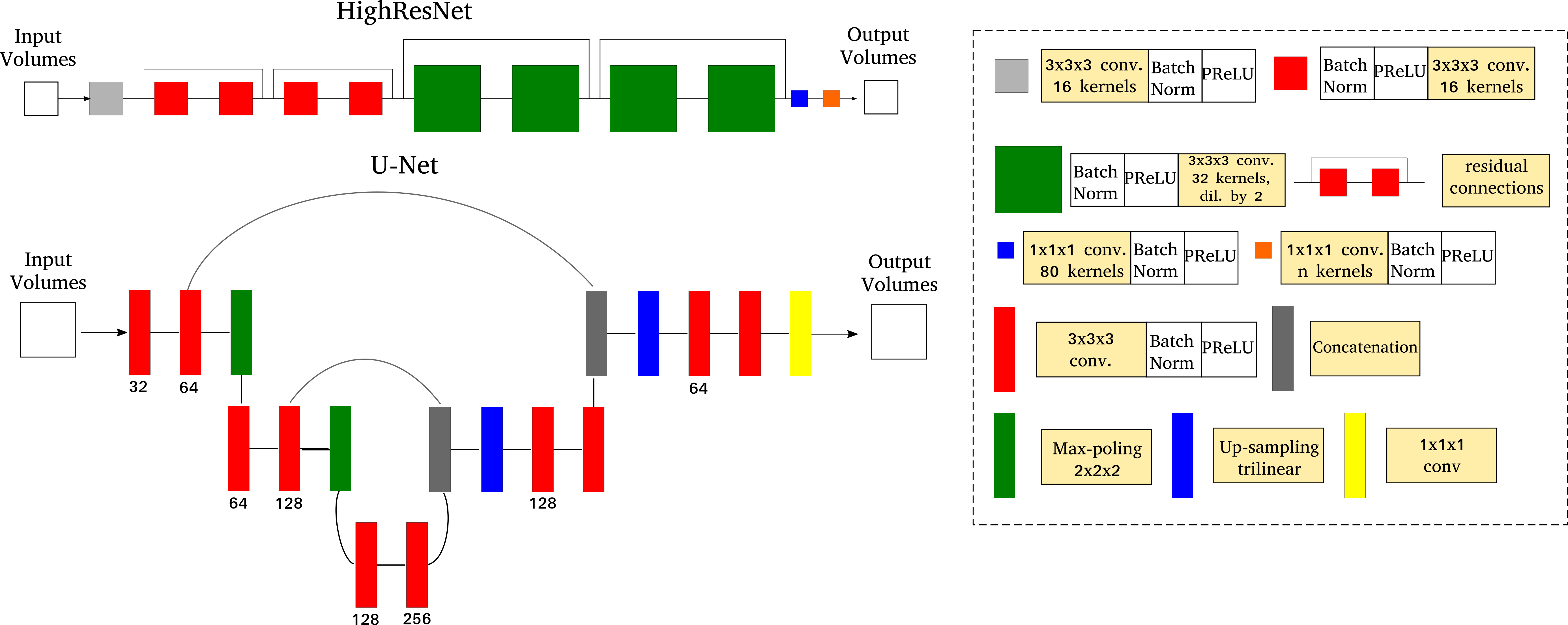}
\caption{\label{fig:networks_pipe} HighResNet architecture~\cite{li2017compactness} and U-Net architecture~\cite{cciccek20163d} used in this work.}
\end{center}
\end{figure*}

We evaluate two 3D CNNs architectures, a 3D High-Resolution Network (HighResNet)~\cite{li2017compactness} and a 3D U-Net~\cite{cciccek20163d} on their ability to regress M-CSD model coefficients from 2TS-CSD model coefficients. Figure~\ref{fig:networks_pipe} shows a graphical representation of the network architectures. For both networks, patch-based training is used making it necessary to reduce the effective receptive field (ERF) to reflect the selected patch size (Section~\ref{sec:training_setup}) to avoid distortions at the patch boundary voxels~\cite{luo2016understanding}.

\subsubsection{HighResNet}
The original HighResNet architecture comprises of $3$ levels of dilated convolutions and $9$ residual connections resulting in $0.81$M trainable parameters. HighResNet was originally proposed as a compact network that could achieve large ERFs~\cite{luo2016understanding} without requiring a downsample-upsample pathway to capture low and high level features~\cite{cciccek20163d,milletari2016v}. Dilated convolutions are used to produce accurate predictions and detailed probabilistic maps near object boundaries~\cite{li2017compactness}.  In this paper, we modified the HighResNet architecture by reducing the number of layers to achieve the desired, reduced ERF. The final HighResNet architecture comprises $2$ levels of dilated convolutions and $4$ residual connections resulting in $0.16$M trainable parameters. A parametric rectified linear unit (PReLU) activation function was used in place of a ReLU as PReLU adaptively learns the rectifier parameters and has been shown to improve CNNs performance in other applications~\cite{he2015delving}. 

\subsubsection{U-Net}
The original 3D U-Net is a “U”-shaped network that has a downsample-upsample pathway composed of $14$ convolutional layers~\cite{cciccek20163d} resulting in $19.08$M trainable parameters. We adapted the U-Net architecture to achieve the desired ERF as the patch size used for training by reducing the network to $10$ convolutional layers resulting in $3.93$M trainable parameters. We removed one encoder block ($2 \times$ (conv. + batch. norm + PReLu) + max pooling) and one decoder block (concat. + up-sampling + $2 \times$ (conv. + batch. norm + PReLu)).

\subsubsection{Data augmentation}
Classical techniques for on-the-fly augmentation including axis flipping, scaling, and rotation have been successfully applied to DL training in small 3D medical imaging datasets~\cite{wasserthal2018tractseg, li2017compactness, gibson2018niftynet}. However, applying these techniques as implemented in traditional medical image processing tools is not appropriate as the CSD coefficients are in the SH domain and not the 3D spatial domain. Therefore, we focused on applying a 3D random rotation in the SH domain to augment our dataset as in~\cite{nath2019deep}. 

\subsection{Training setup}
\label{sec:training_setup}
Each network is trained with an RMSprop optimizer to minimize the $L_{2}$ loss between the M-CSD coefficients and the CSD coefficients regressed by the CNN, measured by $loss\left(y, \hat{y}\right) =  
\frac{\lVert y-\hat{y} \rVert_{2} ^{2}}{2}$ where $y$ are the ground truth coefficients (M-CSD) and $\hat{y}$ are the coefficients regressed by the network. Each CNN is initialized using He uniform function~\cite{he2015delving} and trained for $400$ epochs, based on experimentally chosen convergence, with a weight decay of $1E-6$. Training starts with a learning rate of $3E-2$, which is then reduced by $1/2$ every $50$ epochs. The entire network is trained using patches sampled within intracranial space. Patches were uniformly sampled from within a binary mask corresponding to the intracranial space, computed using the skull-stripped algorithm in MRtrix~\cite{tournier2019mrtrix3}. 

For each iteration in an epoch, a subject from the training set is randomly selected. Subsequently, the dMRI data is augmented by applying a random FOD rotation in the range of $[-25, 25]$ degrees to the data. From this data, $40$ patches of size $32\times32\times32\times 15$ were randomly sampled from within the intracranial space, where $15$ is the number of 2TS-CSD coefficients. The number of patches were experimentally selected to achieve low validation loss while being able to be loaded on the available
graphics processing unit (GPU) memory. An epoch finishes when all subject from the training set have been selected once. For every epoch, a new set of random rotations and patches are computed from each subject.

\label{sec:ExpDesign}
\section{Experimental Design}
\subsection{Datasets}
\label{sec:datasets}
In this work, we use two datasets to conduct our analysis: the publicly available HCP~\cite{sotiropoulos2013advances} and a dataset acquired at the National Hospital for Neurology and Neurosurgery, Queen Square (QS). Each dataset and its preprocessing is found below. 

\subsubsection{QS Dataset} 
The QS dataset is composed of $50$ volumetric MS dMRI scans acquired from patients with epilepsy who appeared ``structurally normal" on a T1-weighted MRI (T1). Small lesions are present focally in the GM but should not distort or affect the WM FOD computation. All patients underwent MRI as part of routine clinical care. Images are acquired on a $3$T GE MR$750$ that included a T1 sequence (MPRAGE) and a MS dMRI sequence with $2$ mm isotropic resolution and the gradient directions $11$, $8$, $32$, and $64$ at b = $0$, $300$, $700$, and $2500$ s/mm$^2$, respectively and single b = $0$ s/mm$^2$ with reverse phase-encoding. QS dataset is corrected for signal drift, geometric distortions and eddy-current
induced distortions as in~\cite{mancini2019automated}.

\subsubsection{HCP Dataset} 
We use a subset of HCP data composed of $45$ subjects~\cite{sotiropoulos2013advances}. Images are acquired on a $3$T scanner with the following parameters: $1.25$ mm isotropic resolution with $90$ gradient directions for each b = $\in$ \{$1000$, $2000$, $3000$ s/mm$^2$\} and $18$ images at b = $0$ s/mm$^2$. Data from the HCP dataset is corrected following the protocols described in~\cite{sotiropoulos2013advances} prior to download.

\subsection{Evaluation Metrics}
To evaluate the accuracy of CSD models we compute mean absolute error (MAE) and the angular correlation coefficient (ACC)~\cite{anderson2005measurement} between the M-CSD coefficients and either CSD coefficients regressed from a CNN or 2TS-CSD coefficients for all voxels in the WM (Section~\ref{sec:csd_modeling}).
The ``ground truth" CSD is the M-CSD computed for data using all gradient directions and shells. 
WM voxels are identified using geodesic information flows (GIF) to segment the WM as a binary mask~\cite{cardoso2015geodesic}

MAE is computed as (function) and is a measure of how well the CSD model coefficients match, lower values are indicative of CSD model coefficients similar to the ground truth CSD model coefficients. MAE is computed as $MAE\left(y, \hat{y}\right) =  
\frac{\lvert y-\hat{y} \rvert}{n}$ where $y$ is M-CSD coefficients and $\hat{y}$ are either the CSD coefficients inferred from a trained network or 2TS-CSD coefficients.

ACC is a similarity metric computed between two different sets of SH coefficients $u$ and $v$ of the same order, where $j$ is the SH order. ACC is computed as:

\begin{equation}
ACC(u,v) = \frac{\sum_{j=1}^{\infty}\sum_{m=-j}^{j} u_{j,m}v_{j,m}^{*}}{\left[ \sum_{j=1}^{\infty}\sum_{m=-j}^{j} u_{j,m}^{2}\right]^{0.5}\left[ \sum_{j=1}^{\infty}\sum_{m=-j}^{j} v_{j,m}^{2}\right]^{0.5} + \alpha}
\end{equation}
ACC has a scale in the interval $[-1,1]$, where $1$ implies a perfect linear correlation between two functions on a sphere, and $-1$ implies a negative correlation~\cite{schilling2018histological}. However, in this work the value of $-1$ is not achievable due to the non-negativity constraint of the CSD models.

\subsection{Experiments}
\label{sec:exps}
We assess the performance of the CNN models in the following scenarios: 1) training and testing on datasets with the same dMRI acquisition protocol (\textit{Intra-Scanner and Acquisition Performance}); 2) testing on datasets with a different dMRI acquisition protocol than the training dataset (\textit{Inter-Scanner and Acquisition Performance}); and 3) testing on datasets with fewer dMRI gradient directions than the training dataset (\textit{Downsampled Imaging Performance}). Additionally, we assess the CNN models with regards variability within specific brain regions. The details of each experiment are described below. For all experiments, the ``ground truth" CSD is the M-CSD computed for data using all gradient directions and shells~(Section~\ref{sec:train_data}).

\subsubsection{Experiment 1 - Intra-Scanner Acquisition Performance}
We assess how well the CNN models were able to regress M-CSD model coefficients when using the same acquisition protocol for training and testing. In this experiment, $5$-fold cross-validation is used where $3$ folds are used for training, $1$ fold for validation, and $1$ fold for testing.

\subsubsection{Experiment 2 - Inter-Scanner Acquisition Performance}
We assess the generalizability of the CNN models to regress CSD models computed from dMRI acquired from a different acquisition protocol. Here, without further model tuning, we use the CNN models trained in Experiment 1 and test on a dataset for a different dMRI acquisition protocols than used during training. For example, a model trained on QS dataset, using b=$700$ s/mm$^2$ 2TS-CSD coefficients as input, will be tested using as input the 2TS-CSD model coefficients computed from the HCP dataset for b=$2000$ s/mm$^2$.

\subsubsection{Experiment 3 - Downsampled Imaging Performance}
\label{sec:experiment3}
We assess robustness of the CNN models when acquisitions have fewer gradient directions than used for training the CNN model. We subsampled the original holdout test data by 25\%, 50\%, and 75\% of the total number of gradient directions.
Models from Experiment 1, with no further tuning, were used to regress M-CSD coefficients from 2TS-CSD coefficients computed with fewer gradient directions than the acquisitions corresponding to the CSD models used to train the networks~(see Section~\ref{sec:datasets} for datasets acquisition details). Similar to Experiment 2, testing on dMRI data with a different acquisition protocol than used during training the CNN models is also performed.

To construct a SS dMRI dataset with fewer gradient directions, we first use the command \textit{dirorder} from MRtrix to reorder the set of gradient directions such that if a scan is terminated prematurely, at any point, the acquired gradient directions will still be close to optimally distributed on the half-sphere~\cite{tournier2019mrtrix3}. Then, we truncated the number of gradient directions for both b=$0$ s/mm$^2$ and the shell selected to generate the SS dMRI to give the required reduction in scans (25\%, 50\%, or 75\%). Finally, the 2TS-CSD is computed for the subsampled SS dMRI.

\subsubsection{Experiment 4 - Variability Within Brain Regions}
\label{sec:brain_parcels}
We assess the performance of the CNN models within specific brain regions. We evaluate how well the model performs for the Corpus Callosum (CC), and the following regions: frontal, occipital, parietal, and temporal lobes. All brain regions were determined from a segmentation generated using geodesic information flows (GIF)~\cite{cardoso2015geodesic} where WM labels are derived from the Hammers Atlas~\cite{hammers2003three}.   


\subsection{Implementation}
All experiments were performed on a workstation equipped with an Intel CPU (Xeon\textregistered~W-2123, $8 \times 3.60$ GHz; Intel), $32$ GB of memory and a NVIDIA GPU (GeForce Titan V) with $12$ GB of on-board memory. All code was implemented in python using PyTorch~\cite{paszke2019pytorch} for the networks training, NiftyNet~\cite{gibson2018niftynet} for data loading and sampling, and SHtools~\cite{wieczorek2018shtools} for SH rotations during data augmentation. The code used to train the CNN models is available online\footnote{https://github.com/OeslleLucena/RegressionFOD}.

\label{sec:Results}
\section{Results}
In this section, we present results from Experiments 1-4 as detailed in Section~\ref{sec:exps}. 
As a reference for Tables and Figures, a method with the acronym QS 700-HCP 2000 \textit{CNN} U-Net indicates that the model was trained on the QS dataset using the 2TS-CSD model coefficients fit from shell b=$700$ s/mm$^2$ and that model is tested on the HCP dataset using 2TS-CSD model coefficients fit from shell b=$2000$ s/mm$^2$. The method called 2TS-CSD (QS 700) indicates the baseline approach was computed on the QS data for the shell b=$700$ s/mm$^2$. The ``ground truth" is always the M-CSD coefficients computed using the full data -- the original number of gradient directions and shells. 

\subsection{Experiment 1}
\label{sec:exp1}
In this experiment, we evaluated how well the CNN models performs when testing on a dataset with the same dMRI acquisition as the training dataset. Table~\ref{tab:acc_exp_1} reports the average MAE and ACC values between the indicated CSD coefficients and the M-CSD coefficients. As shown in Table~\ref{tab:acc_exp_1}, the CSD coefficients from U-Net and HighResNet are the most similar to M-CSD while the baseline 2TS-CSD coefficients are the least similar. Pronounced ACC improvements are found in the QS dataset (an improvement of $9\%$ for the ACC mean) compared to HCP dataset (an improvement of $5\%$ for the ACC mean) for the best CNN models.

\begin{figure*}[!htb]
\begin{center}
\includegraphics[width=\textwidth]{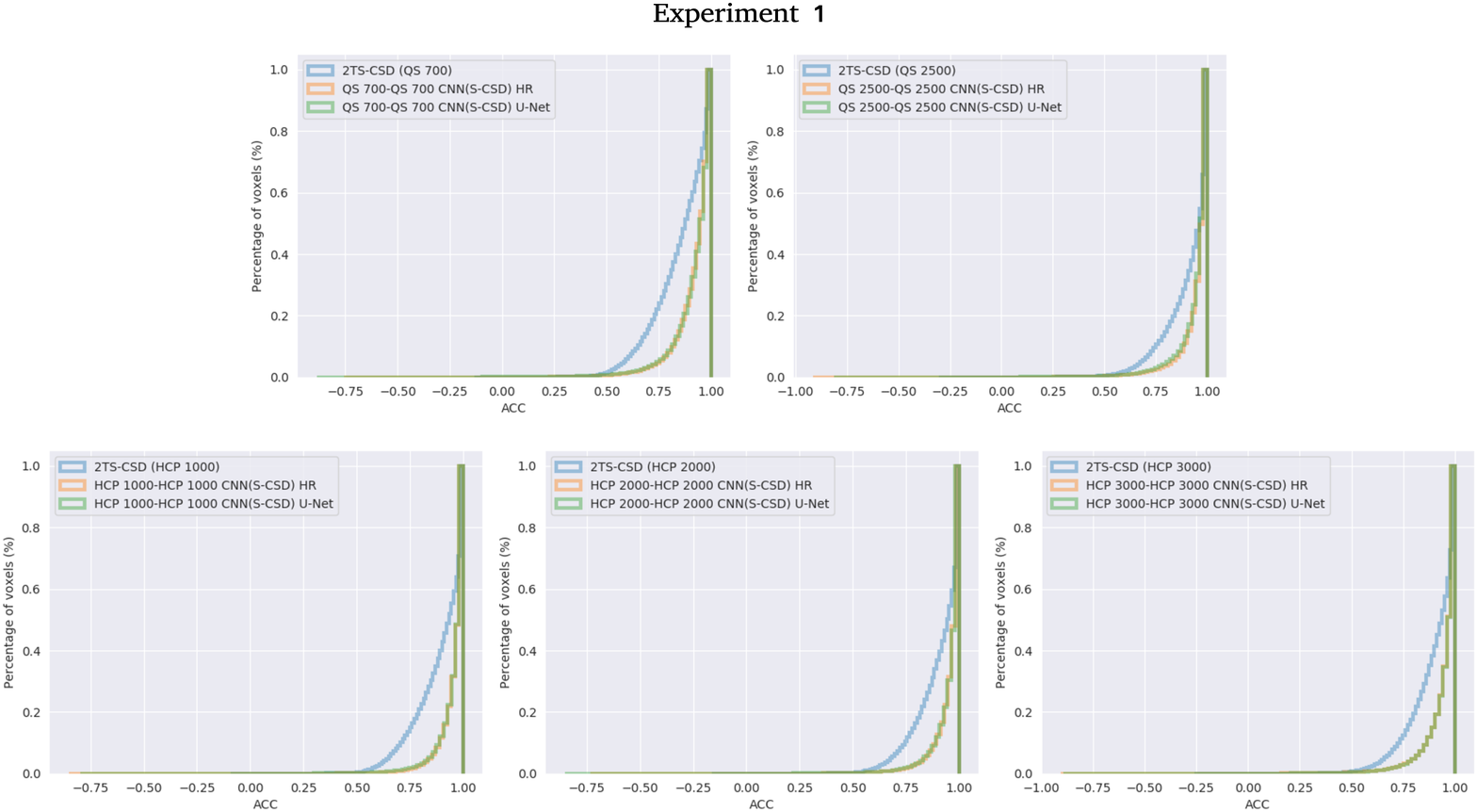}
\end{center}
\caption{\label{fig:acc_cdf_exp1} ACC Cumulative distribution functions (CDFs) forExperiment 1.  2. The ideal ACC would give a single peak at one. HR = HighResNet. Note HR and U-Net have similar CDFs, resulting in an overlay of the green and orange lines.}
\end{figure*}

Figure~\ref{fig:acc_cdf_exp1} shows the cumulative distribution function (CDF) of ACC values over all WM voxels. The CNN models have curves more skewed to high ACC values, demonstrating more voxels with a high level of agreement to the M-CSD compared to 2TS-CSD.

Figures~\ref{fig:acc-hm-qs}~and~\ref{fig:acc-hm-hcp} show qualitative heatmaps of ACC for the WM voxels. Figures~\ref{fig:acc-hm-qs}~and~\ref{fig:acc-hm-hcp}, ACC heatmaps have a high similarity across the three tested methods in WM voxel far from boundaries, where PVE effects are minimal, and low correlation in boundary voxels, which are most likely to have PVE. 

\begin{table*}[!htb]
\centering
\caption{\label{tab:acc_exp_1} MAE and ACC mean, std and median between M-CSD and CSD model coefficients for Experiment 1. The minimum MAE and maximum ACC mean and median are in bold. The label \textit{Train} indicates data used to train the \textit{CNNs} and \textit{Test} indicates the dataset used for the inference. HR = HighResNet.}
\resizebox{\textwidth}{!}{
\begin{tabular}{ccccc}
\toprule
\multirow{2}{*}{\textbf{Train-Test}} & \multirow{2}{*}{\textbf{Method}} 
& \textbf{MAE} 
& \multicolumn{2}{c}{\textbf{ACC}} 
\\\cmidrule{3-5}
&
& mean(std) 
& mean 
& median

\\\midrule
- &  \textit{2TS-CSD} (QS 700) & 0.972(0.079) & 0.851(0.133) & 0.882\\\cdashline{2-5}
\multirow{2}*{QS 700-QS 700} 
& \textit{CNN}  HR & 0.450(0.048) & \textbf{0.928(0.088)} & 0.959\\
& \textit{CNN}  U-Net & \textbf{0.440(0.038)} & 0.927(0.097) & \textbf{0.960}
\\\midrule
- & \textit{2TS-CSD} (QS 2500) & 0.826(0.070) & 0.917(0.107) & 0.966\\\cdashline{2-5}
\multirow{2}*{QS 2500-QS 2500} 
& \textit{CNN}  HR & \textbf{0.307(0.031)} & \textbf{0.958(0.069}) & \textbf{0.981}\\
& \textit{CNN}  U-Net & 0.311(0.030) & 0.954(0.077) & \textbf{0.981}
\\\midrule
- & \textit{2TS-CSD} (HCP 1000) & 0.815(0.130) & 0.895(0.117) & 0.939\\\cdashline{2-5}
\multirow{2}*{HCP 1000-HCP 1000}
& \textit{CNN}  HR & 0.303(0.038) & \textbf{0.961(0.057)} & \textbf{0.983}\\
& \textit{CNN}  U-Net & \textbf{0.298(0.042)} & 0.960(0.063) & \textbf{0.983}
\\\midrule
- &  \textit{2TS-CSD} (HCP 2000) & 0.742(0.123) & 0.909(0.107) & 0.953\\\cdashline{2-5} 
\multirow{2}*{HCP 2000-HCP 2000} 
& \textit{CNN}  HR & 0.284(0.042) & \textbf{0.961(0.062}) & \textbf{0.984}\\
& \textit{CNN}  U-Net & \textbf{0.283(0.038)} & 0.960(0.066) & \textbf{0.984}
\\\midrule
- &  \textit{2TS-CSD} (HCP 3000) & 0.737(0.109) & 0.899(0.112) & 0.941\\\cdashline{2-5}
\multirow{2}*{HCP 3000-HCP 3000}
& \textit{CNN}  HR & 0.310(0.046) & 0.951(0.077) & \textbf{0.980}\\
& \textit{CNN}  U-Net & \textbf{0.303(0.041)} & \textbf{0.952(0.075)} & \textbf{0.980}
\\\bottomrule
\end{tabular}
}
\end{table*}

Finally, Figures~\ref{fig:glyphs_qs} and~\ref{fig:glyphs_hcp} present a visualization of FODs, represented by glyphs showing the direction and distribution of the FOD per voxel. The glyphs show that both U-Net and HighResNet are better able to resolve multiple fiber populations, small rotations and scaling within select regions.


\subsection{Experiment 2}
\label{sec:exp2}
In this experiment, we evaluate how well the CNN models performed when testing on datasets with a different dMRI acquisition protocol than used during training the CNN models compared to the baseline method 2TS-CSD. As models trained with the HCP dataset for different b-values had similar performance in Experiment 1~(Section~\ref{sec:exp1}), we select one model (HCP 2000) for evaluation in Experiment 2.

Table~\ref{tab:acc_exp_2} reports the average MAE and ACC values for the model coefficients. As shown in Table~\ref{tab:acc_exp_2}, both U-Net and HighResNet were quantitatively more similar to M-CSD in terms of ACC and MAE than the baseline 2TS-CSD.

\begin{table*}[!htb]
\centering
\caption{\label{tab:acc_exp_2} MAE and ACC mean, std and median between M-CSD and CSD model coefficients for Experiment 2. The minimum MAE and maximum ACC mean and median are in bold. The label \textit{Train} indicates data used to train the \textit{CNNs} and \textit{Test} indicates the dataset used for the inference. HR = HighResNet.}
\resizebox{\textwidth}{!}{
\begin{tabular}{ccccc}
\toprule
\multirow{2}{*}{\textbf{Train-Test}} & \multirow{2}{*}{\textbf{Method}} 
& \textbf{MAE} 
& \multicolumn{2}{c}{\textbf{ACC}} 
\\\cmidrule{3-5}
&
& mean(std) 
& mean 
& median 
\\\midrule
- &  \textit{2TS-CSD} (QS 700) & 0.972(0.079) & 0.851(0.133) & 0.882\\\cdashline{2-5}
\multirow{2}*{HCP 2000-QS 700}
& \textit{CNN}  HR & \textbf{0.702(0.069)} & \textbf{0.910(0.096)} & \textbf{0.943}\\
& \textit{CNN}  U-Net & 0.702(0.075) & 0.902(0.099) & 0.935
\\\midrule
- &  \textit{2TS-CSD} (QS 2500) & 0.826(0.070) & 0.917(0.107) & 0.966\\\cdashline{2-5}
\multirow{2}*{HCP 2000-QS 2500}
& \textit{CNN}  HR & \textbf{0.428(0.039)} & \textbf{0.945(0.075)} & \textbf{0.971}\\
& \textit{CNN}  U-Net & 0.469(0.034) & 0.938(0.081) & 0.967\\
\midrule
- &  \textit{2TS-CSD} (HCP 2000) & 0.742(0.123) & 0.909(0.107) & 0.953\\\cdashline{2-5}
\multirow{2}*{QS 700-HCP 2000}
& \textit{CNN}  HR & 0.556(0.075) & 0.943(0.070) & 0.971\\
& \textit{CNN}  U-Net & 0.684(0.081) & \textbf{0.946(0.073)} & \textbf{0.973} \\
\cdashline{2-5}
\multirow{2}*{QS 2500-HCP 2000}
& \textit{CNN}  HR & \textbf{0.441(0.079)} & 0.943(0.074) & \textbf{0.973}\\
& \textit{CNN}  U-Net & 0.453(0.088) & 0.939(0.089) & \textbf{0.973}\\
\bottomrule
\end{tabular}
}
\end{table*}

\begin{figure*}[!htb]
\begin{center}
\includegraphics[width=\textwidth]{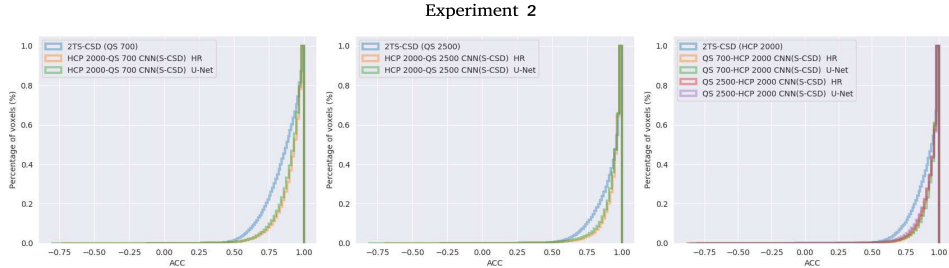}
\end{center}
\caption{\label{fig:acc_cdf_exp2} ACC Cumulative distribution functions (CDFs) forExperiment 2. The ideal ACC would give a single peak at one. HR = HighResNet.}
\end{figure*}

Figure~\ref{fig:acc_cdf_exp2} shows the CDF of ACC values.
Here, we see CNN models have the greatest improvements over the standard 2TS-CSD for the QS 700 data. As in Experiment 1, the CNN models had more skewed CDF curves demonstrating a higher correlation to M-CSD compared to the baseline 2TS-CSD coefficients. 

\begin{figure}[h]
\begin{center}
\includegraphics[width=\textwidth]{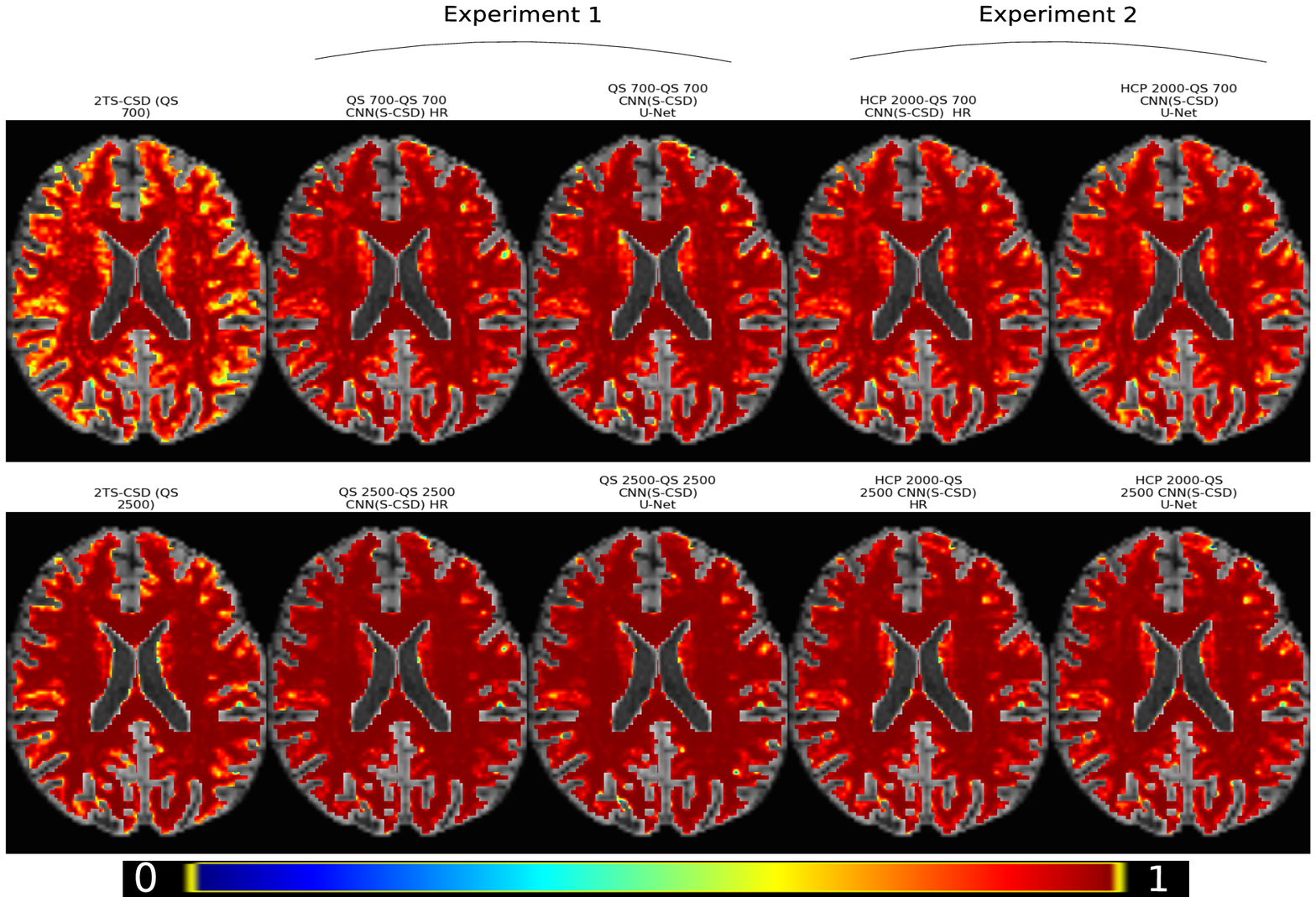}
\caption{\label{fig:acc-hm-qs} ACC heatmaps overlaid on a T1 for one subject from the QS dataset. HR = HighResNet. The ACC values are in the JET colormap and scaled between $[0, 1]$.}
\end{center}
\end{figure}

\begin{figure}[!htb]
\begin{center}
\includegraphics[width=\textwidth]{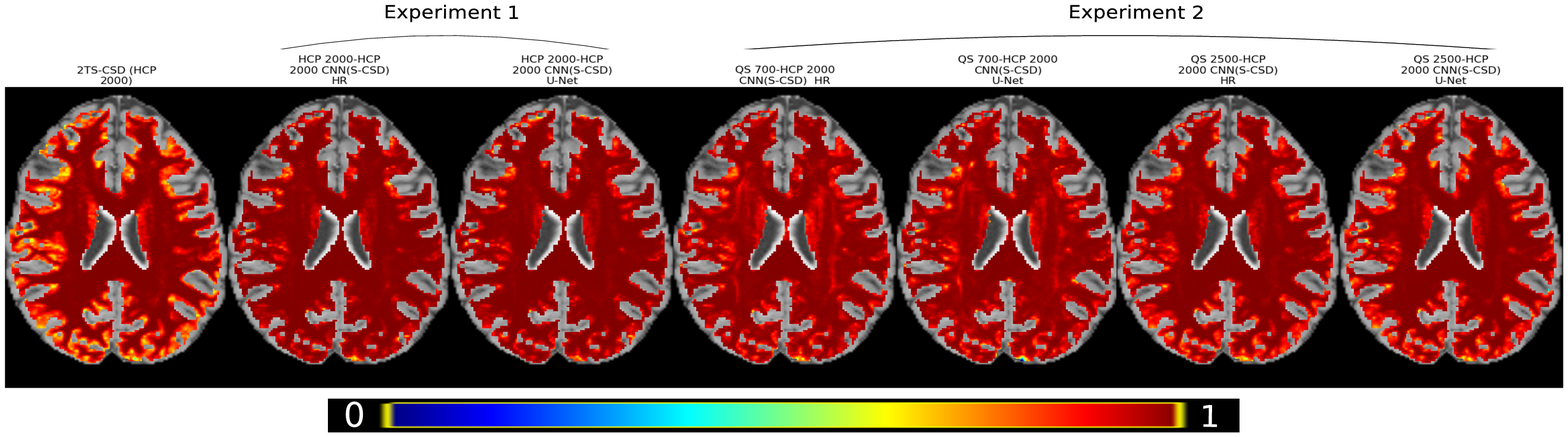}
\caption{\label{fig:acc-hm-hcp} ACC heatmaps overlaid on a T1 for one subject from the HCP dataset. HR = HighResNet. The ACC values are in the JET colormap and scaled between $[0, 1]$.}
\end{center}
\end{figure}

Figures~\ref{fig:acc-hm-qs}~and~\ref{fig:acc-hm-hcp} show heatmaps for ACC for the WM voxels. Although ACC heat maps show higher errors in WM voxels far from the boundary on CNN outputs for Experiment 2 compared to Experiment 1, U-Net and HighResNet still better capture finer details in regions containing fiber crossings and for boundary voxels compared to 2TS-CSD~(Figures~\ref{fig:acc-hm-qs} and~\ref{fig:acc-hm-hcp}). 

\begin{figure}[h]
\begin{center}
\includegraphics[width=\textwidth]{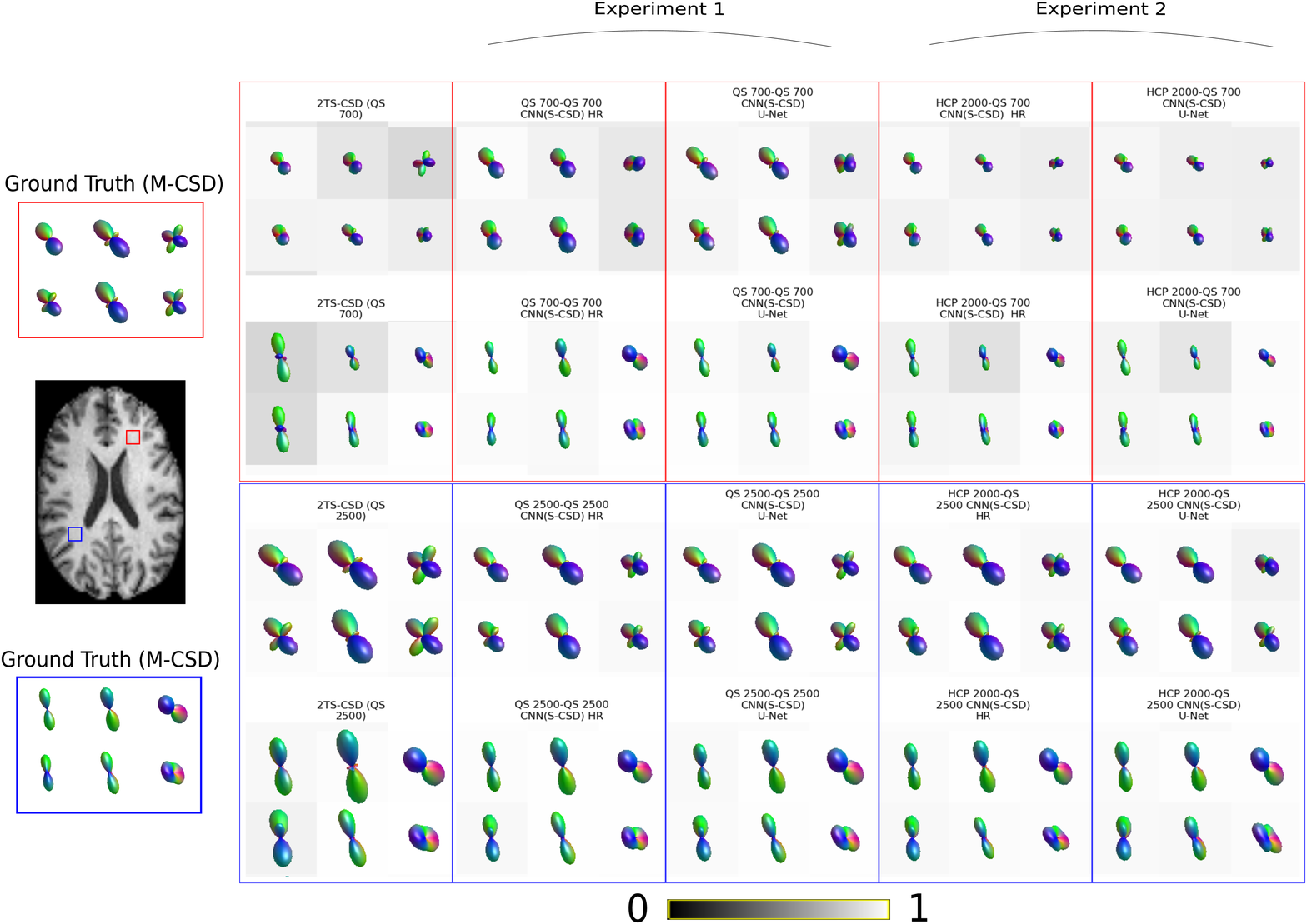}
\caption{\label{fig:glyphs_qs} A glyph representation of the FODs for the regions indicated by the red and blue boxes (magnified by $3\times$) for one subject from the QS dataset. HR = HighResNet. The ACC values are in grayscale and scaled between $[0, 1]$.}
\end{center}
\end{figure}

\begin{figure}[!htb]
\begin{center}
\includegraphics[width=\textwidth]{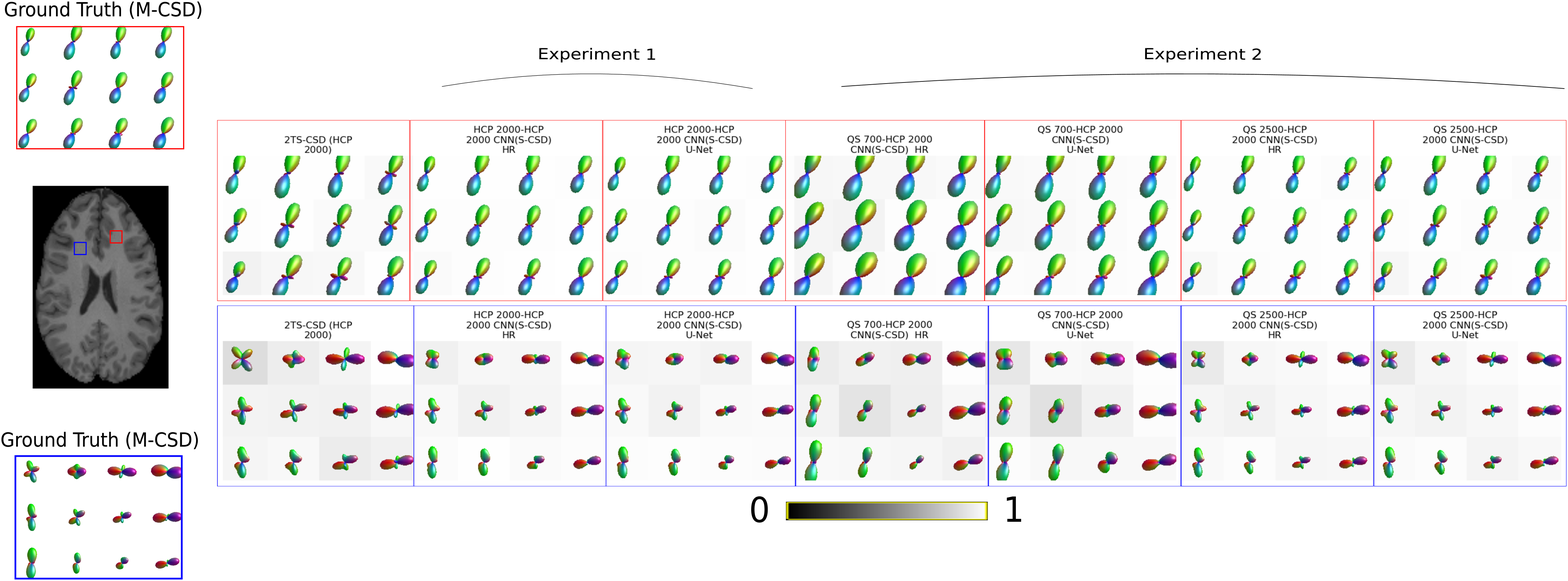}
\caption{\label{fig:glyphs_hcp} A glyph representation of the FODs for the regions indicated by the red and blue boxes (magnified by $1.5\times$) for one subject from the HCP dataset. HR = HighResNet. The ACC values are in grayscale and scaled between $[0, 1]$.}
\end{center}
\end{figure}

Finally, Figures~\ref{fig:glyphs_qs}~and~\ref{fig:glyphs_hcp} show a visual representation of the FODs as glyphs. FOD models for the QS dataset show a higher qualitatively agreement towards single fiber populations compared to multiple fiber populations. Small FODs amplitudes were output  for models trained on the HCP dataset and evaluated on the QS 700 dataset. This may happened due to a poor angular contrast in the low b-value (b=$700$ s/mm$^2$ $32$ directions), which is suboptimal for resolving complex fiber configurations, such that even the CNN model cannot regress the correct CSD coefficients.

As shown in~Figure~\ref{fig:glyphs_hcp}, when regressing the HCP dataset, CNN models trained on QS 2500 dataset were capable of resolving fiber crossings while CNN models trained on the QS 700 dataset could not resolve fiber crossings properly. However, the CNN models trained on QS 700 were able to resolve FOD scaling and small rotations. 

\subsection{Experiment 3}
\label{sec:exp3}
We evaluated how well the CNN models performed when regressing CSD model coefficients of datasets with fewer dMRI gradient directions than used during training. For this experiment, we selected the best CNN models trained for each dataset: b=$2000$ s/mm$^2$ (HCP 2000) and b=$2500$ s/mm$^2$ (QS 2500)~(Section~\ref{sec:exp1}). 

\begin{figure*}[!htb]
\begin{center}
\includegraphics[width=\textwidth]{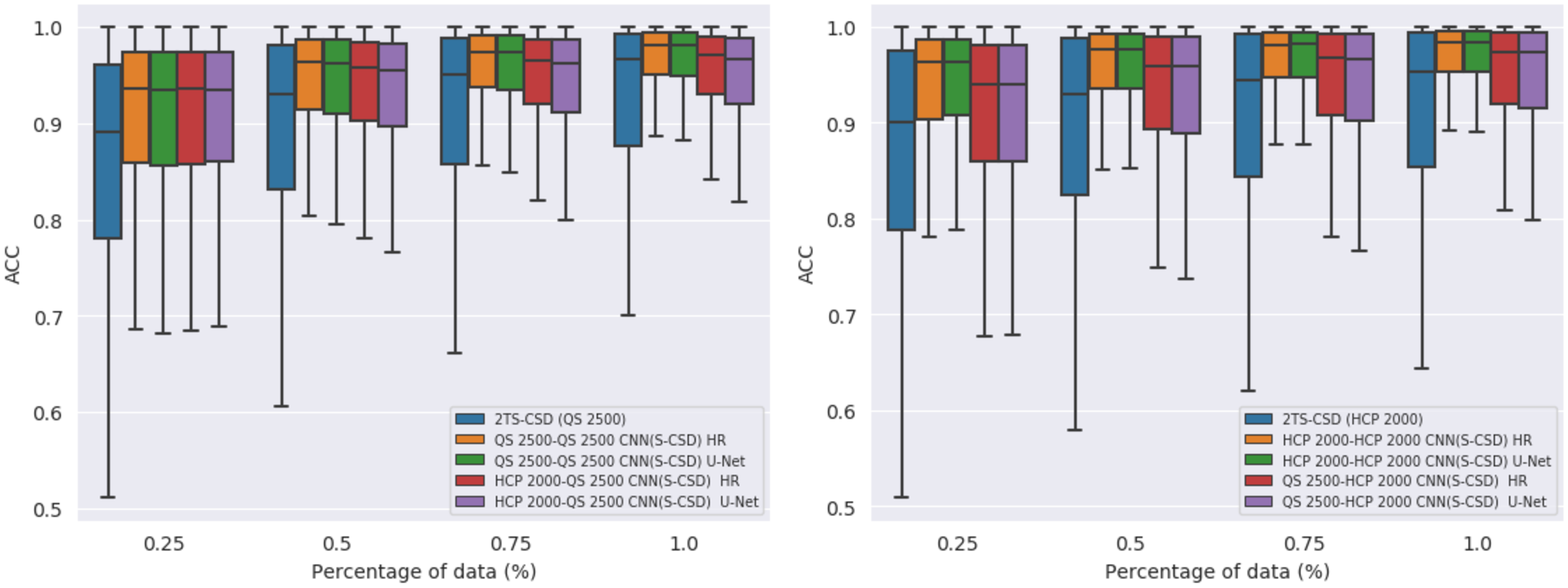}
\caption{\label{fig:subsample-comparison} ACC for different levels of gradient direction subsampling. CNN models on the QS dataset b=$2500$ s/mm$^2$ (left) and the HCP dataset for b=$2000$ s/mm$^2$ (right).}
\end{center}
\end{figure*}

Figure~\ref{fig:subsample-comparison} shows the median, interquartile (IQR) range, minimum, and maximum ACC values for different levels of subsampling of the gradient directions for the test dataset. As shown in Figure~\ref{fig:subsample-comparison}, the CNN models have better performance compared to the baseline for all levels of subsampling, demonstrating robustness in both datasets for a very few gradients when tested on data with same and different dMRI protocols used for training the CNN models (ACC medians $> 0.93$ for $0.25\%$ on QS 2500; ACC medians $> 0.94$ for $0.25\%$ on HCP 2000) when compared to the baseline (ACC medians $< 0.90$ for $0.25\%$ on QS 2500; ACC medians $\approx 0.90$ for $0.25\%$ on HCP 2000). Similar improvements are found when training on the b=$700$ s/mm$^2$ dataset (QS 700) (Supplementary Material Table 1) and its ACC results and qualitative analysis are presented in the Supplementary material Figure 2 and 3.






\subsection{Experiment 4}
\begin{figure*}[h]
\begin{center}
\includegraphics[width=\textwidth]{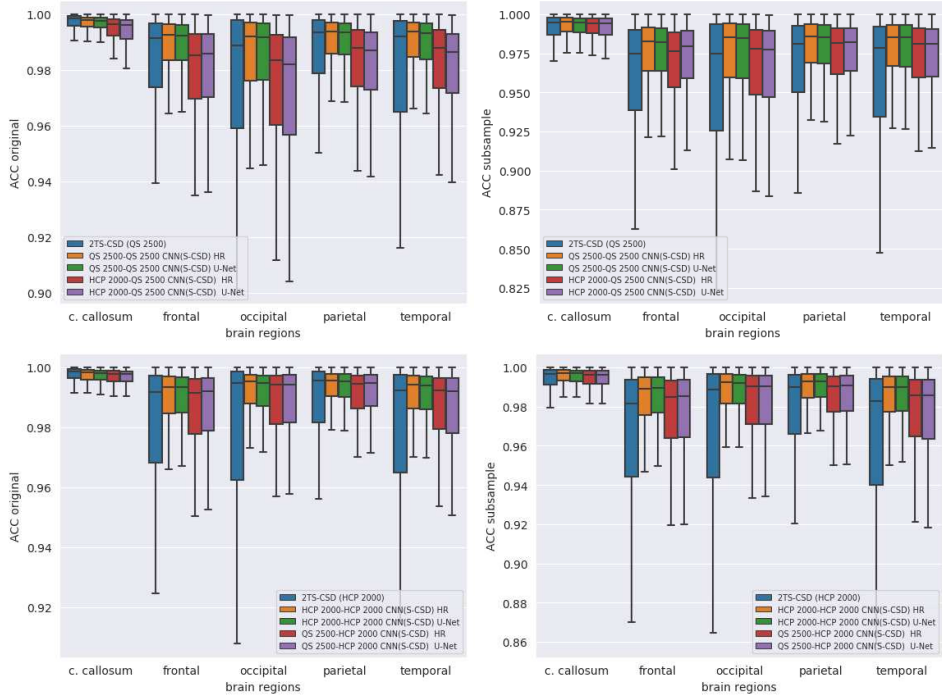}
\caption{\label{fig:brain-parcels} ACC for specific brain regions tested on the QS dataset for b=$2500$ s/mm$^2$ (Top) and the ACC for specific brain regions on HCP dataset for b=$2000$ s/mm$^2$ (Bottom). ACC for specific brain regions when testing with a dataset containing all gradient directions (Left) and with a dataset containing 50\% of the gradient directions as used during training (Right).}
\end{center}
\end{figure*}

In this experiment, we evaluate how well the CNN models regressed CSD coefficients within specific regions of the brain. This experiment can help identifying if there are spatial relationships in how well the CNN models regress the CSD coefficients in the specific brain regions~(Section~\ref{sec:brain_parcels}). Similar to Experiment 3, we selected the best CNN models for each dataset, corresponding to the CNN models QS 2500 and HCP 2000.
Figure~\ref{fig:brain-parcels} shows the ACC for the CC and frontal, occipital, parietal, and temporal lobes. As shown in Figure~\ref{fig:brain-parcels}, ACCs for all brain regions are consistently higher compared to the baseline method (2TS-CSD) when using a model trained and tested on the same dataset.

When testing on HCP dataset, ACC for all brain regions had higher agreement with the ground truth compared to the baseline method when QS 2500 CNN models are used. 
When testing on QS dataset, the HCP 2000 CNN models had good performance in specific regions (occipital and temporal lobes) but relatively poor performance in other regions (CC, frontal and parietal lobes). However, differences between ACC were small (between HCP 2000 CNN model and the baseline 2TS-CSD) and may be the result of random variation within the models.
Additionally, in regions where the 2TS-CSD coefficients are very accurate, the CNN model has a slightly lower performance, reflective of the fact there is very little possible improvement in model accuracy when the ACC is close to 1.
When evaluating the CNN models on their ability to regress M-CSD coefficients on subsampled SS dMRI, ACC has higher agreement with the ground truth compared to 2TS-CSD for all brain regions. This indicates the CNN models for dMRI data are robust to the number of gradient directions acquired in all brain regions.







\label{sec:Discussion}
\section{Discussion}
We evaluated patch-based CNN regression to estimate M-CSD model coefficients from 2TS-CSD model coefficients. We demonstrated quantitatively and qualitatively that the CNN can accurately regress CSD model coefficients using data that are are common in clinical settings - SS dMRI. We showed CNN models can regress M-CSD model coefficients from data with the same dMRI acquisition protocol as the training set (Experiment 1); generalize on dMRI data acquired with a different protocol than the training dataset (Experiment 2); and are robust to dMRI acquisition protocols with fewer gradient directions than the training dataset (Experiment 3). Our results show a CNN can be trained on high-quality models then regress model coefficients from acquisitions with a single shell or fewer gradient directions.

Overall, larger improvements, in terms of MAE and ACC, were observed in a clinical protocol (QS) compared to a high quality research protocol (HCP). The HCP dataset has high spatial and angular resolution, which allows even the 2TS-CSD to resolve complex fiber configurations and compute very accurate FODs compared to the QS dataset, in which individual shells may not have enough gradient directions and overall SNR to capture subtle differences. Because of these factors the CNN regression shows greater improvement estimating CSD coefficients when the 2TS-CSD coefficients are less accurate. 

Our approach may enable faster commercial dMRI acquisition with fewer gradient directions, thereby reducing acquisition times and thus facilitating translation in time-limited clinical environments~\cite{ordonez2019identification}. MS dMRI enables better modeling of FODs, especially with voxels where PVE may complicate estimating the model coefficients~\cite{jeurissen2014multi}.
Our work has the potential to estimate WM CSD model coefficients of similar quality to M-CSD coefficients from SS dMRI data. This method could also be applied to improve the analysis of retrospective data where updating the acquisition is not possible. Additionally, improved estimation of M-CSD coefficients from 2TS-CSD coefficients may benefit tractography~\cite{smith2012anatomically,mancini2019automated}. 

In this work, we investigated CNN-based regression methods between CSD models with $l_{max}=4$ order, comprising $15$ coefficients, to establish a proof of concept. To demonstrate our approach is generalizable to higher order CSD models, we performed additional experiments where we trained the CNNs using CSD models from QS 2500 and HCP 2000 with $l_{max}=8$, comprising $45$ coefficients. This experiment used the same training and testing paradigm as Experiment 1. Supplementary Material Table 1 shows the ACC for the higher order CSD models. The ACC and MAE for the CNN models for $l_{max}=8$ have similar errors to the CNN models evaluated for $l_{max}=4$.




We evaluated two common neural network architectures, U-Net~\cite{cciccek20163d} and HighResNet~\cite{gibson2018niftynet}. Both CNN models perform similarly throughout all experiments. The aim of this work was not to find the best CNN to perform CSD coefficients regression but to show the capability of deep learning to facilitate enhanced CSD models for commercially available dMRI acquisition protocols. One future avenue of research is to investigate how the different networks and trainable parameters influence the networks ability to regress CSD model coefficients.



There are two key limitations in this work. First, we used datasets to train our CNN models using data acquired from the same scanner with the same acquisition protocols. Although we demonstrate that our approaches are capable of generalizing across dMRI acquisition protocols~(Experiments 2 and 3), further improvements in the regression model may be obtained by combining datasets across different sites during CNN models training. Secondly, we did not include a validation on subjects with pathologies that would distort WM tissue connectivity, such as tumors or edema. Although the QS dataset~\cite{mancini2019automated} was acquired from patients with epilepsy, if any small lesions or abnormalities were present they were not big enough to distort normal anatomy. 

\label{sec:Conclusions}
\section{Conclusions}
In this work, we presented a 3D patch-based CNN to regress M-CSD model coefficients from 2T-CSD model coefficients. Two CNN model architectures, U-Net and HighResNet, were evaluated on their ability to regress CSD model coefficients  1) on the same dataset; 2) across different dMRI acquisition protocols, and 3) on dMRI with fewer gradient directions than the training. Our approach may enable robust CSD model estimation on single-shell dMRI acquisition protocols with few gradient directions, allowing in faster dMRI acquisition in clinical settings. Future validation is required to demonstrate this approach generalizes to datasets acquired at multiple sites and on patients with brain pathologies that distort normal anatomy, such as tumors.


\section*{Data availability statement}
\textbf{HCP data} is publicly available dataset\footnote{https://db.humanconnectome.org}. We used the 45 subjects from WU-Minn Retest set. \textbf{QS data} which support the findings in this paper was taken from patients examined as part of routine clinical care. Informed consent was obtained from all individual participants included in the study in accordance with the ethical standards of the institutional and/or national research committee. Consent to publish this data was not obtained and will not be made public to protect patient privacy.

\section*{CRediT authorship contribution statement}
\textbf{Oeslle Lucena:} Conceptualization, Methodology, Software, Data curation, Validation, Formal analysis, Investigation, Writing - original draft, Visualization. 
\textbf{Rachel Sparks:} Conceptualization, Resources,
Writing - review \& editing, Supervision, Project administration, Funding acquisition. \textbf{Sjoerd B. Vos:} Resources, Writing - review \& editing, Supervision. \textbf{Vejay Vakharia:} Resources, Writing - review \& editing, Supervision. \textbf{Seb Ourselin:} Conceptualization, Resources, Writing - review \& editing, Supervision, Project administration, Funding acquisition. \textbf{Keyoumars Ashkan:} Conceptualization, Writing - review \& editing, Supervision, Project administration, Funding acquisition. \textbf{John Duncan:} Resources, Writing - review \& editing, Supervision. 

\section*{Acknowledgments}
This research was funded/supported by the National Institute for Health Research (NIHR) Biomedical Research Centre based at Guy's and St Thomas' NHS Foundation Trust and King's College London and/or the NIHR Clinical Research Facility. Oeslle Lucena is funded by EPSRC Research Council (EPSRC DTP EP/R513064/1). Sjoerd B. Vos is funded by the National Institute for Health Research University College London Hospitals Biomedical Research Centre (NIHR BRC UCLH/UCL High Impact Initiative BW.mn.BRC10269). We also thank NVIDIA for the Titan V GPU used in this work. The views expressed are those of the author(s) and not necessarily those of the NHS, the NIHR or the Department of Health. \textbf{Ethical Approval}. All data were evaluated retrospectively. All studies involving human participants were in accordance with the ethical standards of
the institutional and/or national research committee and with the 1964 Helsinki declaration
and its later amendments or comparable ethical standards. 
\textbf{Conflict of interest.} The authors declare that they have no conflict of interest.






\clearpage
\bibliographystyle{elsarticle/elsarticle-num-names}
\bibliography{ref.bib}







\end{document}